\documentclass[a4paper,12pt,english]{article}

\pdfoutput=1

\usepackage[bindingoffset=0.3cm,textheight=22.5cm,hdivide={2.7cm,*,2.7cm}, vdivide={*,22cm,*}]{geometry}
\usepackage[bookmarksnumbered=true,breaklinks=true]{hyperref}

\usepackage{amsmath,amsfonts,amssymb,babel,slashed,graphicx,color,empheq}

\usepackage[numbers,square,comma,sort&compress]{natbib}

\usepackage[toc,page]{appendix}
\newcommand{\stoptocwriting}{%
  \addtocontents{toc}{\protect\setcounter{tocdepth}{-5}}}
\newcommand{\resumetocwriting}{%
  \addtocontents{toc}{\protect\setcounter{tocdepth}{\arabic{tocdepth}}}}

\makeatletter

\newcommand{\bbm}{\left(\begin{matrix}}
\newcommand{\ebm}{\end{matrix}\right)}
\newcommand{\beq}{\begin{eqnarray}}
\newcommand{\eeq}{\end{eqnarray}}
\makeatother

\newcommand{\del}{\partial}

\newcommand{\be}{\begin{equation}}
\newcommand{\ee}{\end{equation}}

\newcommand{\beqa}{\begin{eqnarray}}
\newcommand{\eeqa}{\end{eqnarray}} \newcommand{\eq}[1]{(\ref{#1})}
\def\nn{\nonumber} \def \bea{\begin{eqnarray}} \def\eea{\end{eqnarray}}
\def\obar{\overline}
\newcommand{\barr}{\begin{array}}
\newcommand{\earr}{\end{array}}
\numberwithin{equation}{section}

\def\a{\alpha}  \def\b{\beta}
 \def\g{\gamma} 
 \def\d{\delta} \def\D{\Delta}
    \def\k{\kappa}
 \def\L{\Lambda}  
    
  \def\t{\tau}

\def\cA{{\cal A}}  \def\cC{{\cal C}} 
   
 \def\cH{{\cal H}}  
   
\def\cM{{\cal M}} \def\cN{{\cal N}}

\def\R{{\mathbb R}} \def\C{{\mathbb C}} 

 \def\one{\mbox{1 \kern-.59em {\rm l}}}

\def\msu{\mathfrak{su}}
\def\mso{\mathfrak{so}}

\def\bit{\begin{itemize}} \def\eit{\end{itemize}} \def\Tr{\mbox{Tr}}

\def\({\left(} \def\){\right)}

\newcommand{\diag}{\rm diag}

 \allowdisplaybreaks[3]

\begin{document}

\makeatother


\parindent=0cm

\renewcommand{\title}[1]{\vspace{10mm}\noindent{\Large{\bf

#1}}\vspace{8mm}} \newcommand{\authors}[1]{\noindent{\large

#1}\vspace{5mm}} \newcommand{\address}[1]{{\itshape #1\vspace{2mm}}}


\begin{titlepage}
\begin{flushright}
 UWThPh-2017-31 
\end{flushright}
\begin{center}
\title{ {\Large  Cosmological space-times with
resolved Big Bang  \\[1ex]
in Yang-Mills matrix models}  }

\vskip 3mm

\authors{Harold C. Steinacker}

\vskip 3mm

 \address{ 

{\it Faculty of Physics, University of Vienna\\
Boltzmanngasse 5, A-1090 Vienna, Austria  }  
  }

\bigskip

\vskip 1.4cm

\textbf{Abstract}
\vskip 3mm

\begin{minipage}{14cm}%

We present simple solutions of IKKT-type matrix models 
that can be viewed as quantized  
homogeneous and isotropic cosmological space-times, with finite density of microstates and a regular Big Bang (BB).
The BB arises from a signature change of the effective metric 
on a fuzzy brane embedded in Lorentzian target space, in the presence of a quantized 4-volume form.
The Hubble parameter is singular at the BB, and becomes small at late times. 
There is no singularity from the target space point of view, and the brane 
is Euclidean ``before'' the BB.
Both recollapsing and expanding universe solutions are obtained, depending on the mass parameters.

\end{minipage}

\end{center}

\end{titlepage}

\tableofcontents

\section{Introduction}

The evolution of the universe and its origin in a Big Bang (BB) appear to be well described by the 
$\L$CDM model of inflationary cosmology. This model is based on general relativity (GR), assuming 
suitable matter content and initial conditions. Nevertheless, the situation is not satisfactory.
The model requires
a dominant role of unknown matter and energy, while postulating that 
GR still applies at cosmological scales.
At very short distances, GR quite certainly breaks down, and a quantum theory of gravity must take over.
This is essential to address the local and global singularities of space-time, in particular the BB, but it also 
leads to serious fine-tuning problems.

There are strong reasons to expect that in a consistent quantum theory including  gravity,
there should be only finitely many degrees of freedom per unit ``volume''.
While we do not know the correct micro-structure of space-time,  
it requires a pre-geometric origin of space-time. This also
seems to be the most reasonable way to  resolve  the singularities in black holes and the BB. 

Among the many possible approaches to this issue, we will follow  an approach based on 
matrix models. By their very nature as discrete pre-geometric models, 
they provide natural candidates to address the above issues. 
Among all pure matrix models, the IKKT model  \cite{Ishibashi:1996xs} is singled out by virtue of maximal 
supersymmetry, and it was proposed as a candidate for a non-perturbative description of IIB string theory.
Although there is at present no solid understanding of this model at a non-perturbative, background-independent level,
there is a good picture of branes arising as classical solutions, with IIB supergravity interactions arising 
at the loop level \cite{Ishibashi:1996xs,Chepelev:1997av,Steinacker:2016nsc}. 
The effective geometry of such branes (given by some matrix background\footnote{Note that the  branes  
should be viewed as classical condensates here, 
this is {\em not} a holographic scenario, and it does not rely on quantum effects.}) can  be elaborated 
as noncommutative or semi-classical geometry, and fluctuations of these backgrounds lead to noncommutative
gauge theory coupled to this geometry \cite{Aoki:1999vr,Szabo:2001kg}. 
Here the maximal supersymmetry of the matrix model plays an important role, since otherwise 
 unacceptable large non-local effects due to UV/IR mixing \cite{Minwalla:1999px,Steinacker:2016nsc})
 invalidate the semi-classical picture.


In this paper, we will present explicit and simple brane solutions of the IKKT matrix model 
with mass term, which can 
serve as (toy-) models for cosmological space-times, and exhibit a BB-like singularity.
They have a space-like $SO(4)$ isometry, and reduce to homogeneous and isotropic FRW cosmologies with $k=1$
in the semi-classical limit. 
These solutions are obtained from basic quantized (``fuzzy'') homogeneous spaces, specifically 
the fuzzy 4-sphere $S^4_N$ and the fuzzy 4-hyperboloid $H^4_n$. These turn out to be solutions of the 
{\em Lorentzian} matrix model in the presence of suitable mass terms, which are different for 
the space-like and time-like matrices. 
The BB  arises from a signature change in the effective metric, {\em taking into account
the quantized 4-volume form} which arises from the 
non-commutative structure of the brane.
The point is that the effective metric on the brane $\cM$
is not the induced metric, but
involves the Poisson structure on the brane in an essential 
way\footnote{The effective metric can be thought of as open string metric 
in the Seiberg-Witten limit \cite{Seiberg:1999vs}.} 
\cite{Steinacker:2016vgf,Steinacker:2010rh,Seiberg:1999vs}. 
The Poisson structure gives rise to the frame bundle, 
and its flux provides the measure for the integration on $\cM$. 
This  determines the conformal factor of the metric
which is singular at the location of signature change, leading to a singular
initial expansion.

It is well-known that fuzzy spaces can be solutions of Lorentzian matrix models, cf.
\cite{Steinacker:2011wb,Kim:2012mw,Jurman:2013ota,Chaney:2015ktw}.
Even compact solutions were found in \cite{Chaney:2015mfa}, 
where it was pointed out that the induced metric on the brane can change from Euclidean to Minkowski signature.
However, this alone is not sufficient to obtain a Big Bang, and it does not imply a rapid expansion. 
The present work differs from the previous ones in two important ways. 
First, we obtain 3+1-dimensional space-time solutions which 
are completely homogeneous and isotropic; more precisely, they
are  covariant under  $SO(4)$ acting on the spatial $S^3$.
Second and most remarkably, a BB with rapid (singular) initial expansion is shown to arise automatically
on these solutions. 
These space-times are governed not by GR but by the matrix model.


We provide two basic examples of such cosmological matrix space-times with BB, one describing a 
recollapsing universe with a big crunch, and one which is expanding forever. 
Although neither seems to agree very well with the standard cosmology 
(at least under the  present crude analysis),
they illustrate how such quantum space-times might look like, and provide a possible {\em explanation} of the BB, 
beyond postulating that it arises from random quantum fluctuations as in other approaches \cite{Vilenkin:1982de}. 
The BB here is simply a feature of the emergent geometry, which is extended by a Euclidean regime.
It arises in the presence of different space-like and time-like masses 
$m^2 \neq m_0^2$  in the matrix model action satisfying certain conditions. 
Even though the solutions may not be realistic and 
stability at the quantum level is not established, they
nicely illustrate the appeal and the scope of the IKKT model 
(or similar matrix models) as a fundamental theory of space-time and matter.

%

\section{Lorentzian matrix models}
\label{sec:MM}

 We are interested in solutions of the following IKKT-type matrix model \cite{Ishibashi:1996xs} with mass terms 
\begin{align}
 S[Y,\Psi] &= \frac 1{g^2}\Tr \Big([Y^a,Y^b][Y^{a'},Y^{b'}] \eta_{aa'} \eta_{bb'} \, - m^2 Y^i Y^i + m_0^2 Y^0 Y^0 
  + \obar\Psi \Gamma_a[Y^a,\Psi] \Big) \ . 
 \label{bosonic-action}
\end{align}
Here $\eta_{ab} = diag(-1,1,...,1)$ 
is interpreted as Minkowski metric of the target space $\R^{1,D-1}$.
Indices $i$ indicate Euclidean directions, and $0$ is  the time-like direction.
Fermions $\Psi$ are included via the Gamma matrices 
$\Gamma^a$ to enable supersymmetry, however we will focus on the bosonic sector from now on.
The above model leads to the classical equations of motion 
\begin{align}
 -\Box_Y Y^i - m^2 Y^i = 0  \nn\\
 \Box_Y Y^0 + m_0^2 Y^0 = 0
 \label{eom-lorentzian-M}
\end{align}
where 
\begin{align}
  \Box_Y = \eta_{ab} [Y^a,[Y^b,.]] 
  \label{Box-Y}
\end{align}
plays the role of the d'Alembertian. We will study solutions of these equations which are  
interpreted as 3+1-dimensional  space-times, more specifically as noncommutative ``branes`` 
embedded in target space.

As emphasized in the introduction, the choice of the matrix model is important. 
The picture of classical brane solutions $\cM$ is presumably justified only for the 
 maximally supersymmetric IKKT model  with $D=9+1$ \cite{Ishibashi:1996xs}, due to UV cancellations of the 
quantum fluctuations; in fact this model reduces to $\cN=4$ SYM on 
4-dimensional backgrounds\footnote{This is the only model of a noncommutative gauge theory
 where quantum corrections are tame and expected to be perturbatively finite \cite{Hanada:2014ima}.}.
 Thus although the solutions given below are not supersymmetric, the underlying 
model \eq{bosonic-action} is, up to the soft mass terms. Hence SUSY is broken spontaneously and softly, but we expect 
that this still ensures sufficient UV cancellations to tame the quantum corrections.
%

These mass terms are important because they introduce a scale into the model, and conversely 
quantum corrections are expected to induce such mass terms on curved backgrounds.
Indeed as discussed in \cite{Kim:2012mw}, after taking into account an IR cutoff and integrating out the scale factor
in the matrix path integral 
\begin{align}
 Z = \int dYd\psi e^{i S_{\rm IKKT}[Y,\psi]}
 \label{path-integral}
\end{align}
the equations of motions \eq{eom-lorentzian-M} arise, with $m^2 \neq m_0^2$ resulting from an 
IR regularization which mildly breaks Lorentz invariance. 
Since we only study classical 
solutions of \eq{eom-lorentzian-M}  and their  geometrical properties, we will restrict ourselves to 
the bosonic part of \eq{bosonic-action}, including the mass terms by hand.
Moreover, we will see that  a Big Bang arises from the present solutions only if $m^2 \neq m_0^2$.
In fact there are no finite-dimensional non-trivial solutions without  a mass term,
as shown in the appendix.


\section{Recollapsing universe from  fuzzy 4-spheres}
\label{sec:recollaps}

\subsection{The Euclidean fuzzy 4-sphere}
\label{sec:fuzzy-S4}

We briefly recall the definition of fuzzy 4-spheres \cite{Grosse:1996mz}, cf. \cite{Castelino:1997rv,Ramgoolam:2001zx,Kimura:2002nq,Medina:2002pc}.
The starting point is the Lie algebra $\mso(6) \cong \msu(4)$, with generators
$\cM^{ab}, \ a,b=1,...,6$ and commutation relations
\begin{align}
  [\cM_{ab},\cM_{cd}] &=i(\d_{ac}\cM_{bd} - \d_{ad}\cM_{bc} - \d_{bc}\cM_{ad} + \d_{bd}\cM_{ac}) \ .
 \label{M-M-relations}
\end{align}
Now consider the embedding of $SO(5) \subset SO(6)$ defined by restricting the indices of 
$\cM^{ab}$ to be in $\{1,...,5\}$, and denote the remaining generators as
\begin{align}
 X^a &= r \cM^{a6}, \qquad a = 1,...,5 \ , \nn\\
  [X^a,X^b]  &= \Theta^{ab} =   r^2 \cM^{ab}
 \label{X-def}
\end{align}
Here $r$ is a scale with dimension length.
By construction, the $X^a$ transform covariantly under
$SO(5)$ generated by $\cM^{ab}$,
\begin{align}
 [\cM^{a b},X^c] &= i(g^{a c} X^b - g^{b c} X^a),
\end{align}
We fix the $SO(6)$ representation to be $\cH = (0,0,N) = (\C^4)^{\otimes_S N}$, which
is well-known to remain irreducible under $SO(5)$. 
Therefore the radius is a constant,
\begin{align}
 X^a X_a &= X^a X^b \d_{ab} = R^2 \one = r^2 R_N^2 \one, \qquad  R_N^2 = \frac 14 N(N+4) \ .
 \label{euclid-sphere}
\end{align}
The $\mso(6) \cong \msu(4)$ generators $\cM^{ab} \in End(\cH), \ a,b=1,...,6$ are now understood as 
quantized embedding functions 
\begin{align}
 \cM^{a b} \sim m^{ab}: \quad \C P^3 \hookrightarrow \mso(6) \ \cong \ \R^{15}
\end{align}
where $m^{ab} = r^{-2}\theta^{ab}$, and similarly 
\begin{align}
 X^{a} \sim x^{a}: \quad \C P^3 \hookrightarrow \R^5 \ .
 \label{X-S4-embed}
\end{align}
In the semi-classical limit, the commutators reduce to the Poisson bracket on $\C P^3$, and 
we can work with the Poisson structure 
\begin{align}
 \{x^a, x^b\} = i \theta^{ab} \ .
\end{align} 
Therefore the semi-classical geometry underlying fuzzy $S^4_N$ is $\C P^3$, which is an $S^2-$ bundle over $S^4$
carrying a canonical symplectic structure, and $x^a:\ \C P^3 \to S^4 \subset \R^5$ is nothing but the Hopf map.  
This can also be justified e.g. via coherent states $|x,\xi\rangle$, 
which are in one-to-one correspondence 
(up to a phase) to points on $\C P^3\cong SU(4)/SU(3)\times U(1)$, which is locally
isomorphic to $S^4 \times S^2 \ni (x,\xi)$.
It turns out that $\theta^{ab} = \theta^{ab}(x,\xi)$ is tangential $x_a \theta^{ab} = 0$ on $S^4$,
and transforms under the local stabilizer $SO(4)_x$ of any point $x \in S^4$.
More precisely, it forms a  bundle of self-dual bi-vectors $\theta^{\mu\nu}$ 
on $S^4$, which is locally isomorphic to $S^4 \times S^2$.
In particular, $[\theta^{ab}]_{S^2} = 0$ 
where $[.]_{S^2}$ denotes the 
 averaging over the internal $S^2$. 
 For more details on fuzzy $S^4_N$ we refer to \cite{Steinacker:2015dra,Sperling:2017dts,Medina:2002pc}. A gentle introduction 
 to the geometrical concepts of fuzzy spaces can be found e.g. in \cite{Steinacker:2011ix}.

\subsection{Lorentzian fuzzy 4-sphere in $\R^{1,4}$}
\label{sec:lorentzian-S4}

We will show that ellipsoidal
deformations of $S^4_N$ are exact solutions\footnote{It is well-known that fuzzy $S^4_N$ is a solution upon 
including a quintic term $Tr (\varepsilon YYYYY)$ \cite{Kimura:2002nq}. However this is not a soft term, and
thus quantum effects are problematic \cite{Azuma:2004yg}. 
Here we show that such a term is not necessary in the presence of a mass term. The 4-dimensional cosmologies in 
\cite{Chaney:2015ktw} are not fully covariant but carry Poisson-structures which break $SO(4)$ invariance. 
This is avoided here.} 
of the model \eq{bosonic-action}, provided
$m^2 \neq m_0^2$. 
Thus let $\cM^{ab}$ be hermitian generators of an irrep of $SO(6)$  as above which remains 
irreducible under $SO(5)$.
Define  $Y^a, \ a\in\{0,...,4\}$ by
\begin{align}
    Y^i &=  X^i , \quad \mbox{for} \ \  i = 1,...,4, \qquad  Y^0 = \k X^5
 \label{S4-ansatz}
\end{align}
for $X^a = r \cM^{a 6}$ as in \eq{X-def}.
Clearly the $Y^i, i=1,...,4$ transform as vectors under $SO(4)\subset SO(5)$.
We ask these $Y^a$  to be solutions of the mass-deformed matrix model \eq{bosonic-action},
which in terms of the $X^a$ variables looks as follows
\begin{align}
 S &= \frac 1{g^2}\Tr \Big([X^a,X^b][X^{a'},X^{b'}] g_{aa'} g_{bb'} \, 
 - m^2  X^i X^i + \tilde m_0^2 X^0 X^0  \Big) \ ,
 \label{bosonic-action-X}
\end{align}
Now the target space metric in these coordinates is
\begin{align}
g_{ab} = \diag(-\k^2,1,1,1,1) \ , \qquad \mbox{and} \ \ \ \tilde m_0^2 = \k^2 m_0^2 \ .
\label{eta-tilde}
\end{align}
The commutation relations \eq{X-def} give
\begin{align}
  [X^a,[X^a,X^b]] &= i r^2 [X^a,\cM^{a b}] = i r^2 [\cM^{ba},X^a] \qquad\mbox{(no sum)}  \nn\\
    &= r^2 \left\{\begin{array}{ll}
                X^b, & b\neq a  \\
               0, & b = a
              \end{array}\right. \ .
\end{align}
Hence the equations of motion
\begin{align}
 (\Box_X + \tilde m_0^2) X^0 = 0 = (\Box_X +m^2) X^i 
\end{align}
 imply 
\begin{align}
 4 r^2 + \tilde m_0^2 = 0 = (3-\k^2)r^2  + m^2 \ .
\end{align}
This clearly requires $m_0^2 < 0$, and
\begin{align}
 \frac{3-\k^2}4 = \frac{m^2}{\tilde m_0^2} = \frac{m^2}{\k^2 m_0^2} \ .
\end{align}
Hence for any $m_0^2 < 0 \leq m^2$ in the original model \eq{bosonic-action}, there is a unique 
\begin{align}
\k^2 \geq 3, \qquad \mbox{or} \ \ 
\k^2 = 3 \quad \mbox{for} \quad  m=0
\label{k-bound-S4}
\end{align}
and $r^2 > 0$ so that $Y^a$ is a solution of \eq{bosonic-action}, or equivalently
$S^4_N$ with  $X^a X^b \d_{ab} = R^2$ \eq{euclid-sphere} is a solution 
of the  matrix model with Lorentzian target space metric   
$g_{ab}$ \eq{eta-tilde}.
There are also $S^4_N$ solutions for $m_0^2<m^2<0$ as long as $\frac{m^2}{m_0^2} \leq \frac{9}{16}$, but 
we will see that they do not acquire a Minkowski metric. However
for $m^2 < m_0^2 < 0$, we will find expanding universe solutions with Minkowski metric, which are 
discussed in section \ref{sec:open-universe}.

To study the geometry in more detail, we restrict ourselves to the semi-classical limit from now on, 
replacing commutators by Poisson brackets as discussed in section \ref{sec:fuzzy-S4}. 
Then \eq{X-S4-embed} is replaced by $X^a \sim x^a:\ \C P^3 \hookrightarrow S^4 \subset \R^{1,4}$.
Hence the image of $\C P^3$ in $\R^{1,4}$ defines a manifold  $\cM$
which is topologically  a 4-sphere carrying a bundle of  bivectors
$\theta^{\mu\nu}$ (which are self-dual w.r.t. its Euclidean $SO(5)$-invariant metric), 
but  embedded in Lorentzian 
target space $\R^{1,4}$. All these structures will play a role, and one must be careful to 
use them appropriately.

We are particularly interested in the metric on $\cM$. 
There are in fact two different metrics on the brane $\cM\subset \R^{1,4}$, as in string theory: 
The {\em induced metric} is simply the pull-back of the constant 
(''closed string``)  metric in target space  $\R^{1,4}$, and it will be determined first.
This is distinct from the {\em effective metric}, which governs the
(noncommutative) gauge theory 
{\em on} the brane $\cM$, which arises from fluctuations\footnote{Note that $\cA^a \in End(\cH) \cong \cC(\C P^3)$
describes indeed functions living {\em on} $\cM$.} $X^a \to X^a + \cA^a$ in the matrix model.
This is the analog of the open string metric \cite{Seiberg:1999vs}, and it will be determined 
in a second step. For a more general discussion of these topics see e.g. \cite{Steinacker:2010rh}.

\paragraph{Induced metric.}

As a warm-up,
we compute  the induced metric $g_{\mu\nu}$ on $\cM \subset \R^{1,4}$. This clearly has Euclidean signature 
at $x_0 = \pm R$, and Minkowski signature for $x_0 \approx 0$. The  domains of fixed signature are separated
by a space-like $S^3\subset S^4$ where the 
metric is degenerate. This is the locus on $S^4$ where the tangent space includes a null
direction of $\R^{4,1}$.
Using the  space-like $SO(4)$ symmetry,
we can choose a standard reference point 
\begin{align}
 x=(x_0,x_1,0,0,0) = R(\cos(\eta),\sin(\eta),0,0,0) \ \in S^4 \subset \R^{4,1} , \qquad x_0^2 + x_1^2 = R^2 \ 
 \label{ref-point}
\end{align}
and use the tangential coordinate 
\begin{align}
 \tau = R \eta
\end{align}
which points in the $x^0 x^1$ direction,  
\begin{align}
 \frac{d}{d\t} x = R(-\sin(\eta),\cos(\eta),0,0,0) \ . 
\end{align}
Then the induced metric is 
\begin{align}
 g_{\mu\nu} = \diag(\cos^2(\eta) - \k^2\sin^2(\eta),1,1,1)
 \label{g-cartesian}
\end{align}
in local 
$x^\mu = (\t x^2 x^3 x^4)$ coordinates on $T_p\cM$ at the standard reference point \eq{ref-point},
or
\begin{align}
 ds^2_g &= R^2 (\cos^2(\eta) - \k^2\sin^2(\eta))d\eta^2 + R^2\sin^2(\eta) d\Omega_3^2  \nn\\
  &= \b^2(\eta) d\eta^2 + \a^2(\eta)d\Omega_3^2 
\end{align}
in FRW coordinates where $d\Omega_3^2$ is the $SO(3)$ -invariant metric on the unit sphere $S^3$ and
\begin{align}
 \b^2(\eta) =  \frac 12 R^2 \big((\k^2+1)\cos(2\eta) -(\k^2-1)\big), \qquad \a^2 = \frac 12 R^2(1-\cos(2\eta)).
\end{align}
Clearly $\a\geq 0$ vanishes only on the poles $x^0 = \pm R$ where $\eta=0,\pi$. In contrast, 
$\b(\eta_*) = 0$ vanishes if
\begin{align}
 \cos(2\eta_*) = \frac{\k^2-1}{\k^2+1} \ , 
 \label{eta-star}
\end{align}
which  is $\eta_* = \frac {\pi}6$ for $\k^2=3$.
Hence  there is indeed an interesting transition from Euclidean to Minkowski signature, however
the associated singularity cannot be interpreted as Big Bang, as there is  no rapid initial expansion.
Therefore the induced metric does not give rise to an interesting cosmology.
In contrast,
we will see that a Big Bang does arise for the effective metric. 
The volume-form arising from the  4-form flux will be crucial for this mechanism.

\paragraph{Effective metric and averaging.}

We now compute the effective metric 
on $\cM \subset \R^{1,4}$ in the matrix model.
It is easiest to use the $x^a$ description where the embedding is spherical, but the target space metric is  
$g_{ab}$ \eq{eta-tilde}.
The {\em effective metric} $G^{\mu\nu}$ in matrix models is determined by the kinetic term 
for a scalar field\footnote{The conformal factor cannot be determined from gauge fields 
because of conformal invariance.} as follows \cite{Steinacker:2010rh,Steinacker:2016vgf,Steinacker:2011ix}
\begin{align}
  S[\phi] &= -  \Tr [X^a,\phi][X^b,\phi] g_{ab} 
  \ \sim \ \frac {\dim \cH}{\rm Vol_\omega(\cM)}\int_{\cM}d^4 x\, \sqrt{|\theta^{\mu\nu}|^{-1}}\, 
  \g^{\mu\nu}\del_\mu \phi \del_\nu \phi \ \nn\\
   &= \ \int_\cM d^4 x\, \sqrt{|G_{\mu\nu}|}\,
  G^{\mu\nu}\del_\mu \varphi \del_\nu \varphi \ .
  \label{kinetic-term-metric}
\end{align} 
using greek indices for local coordinates on $\cM=S^4$.
Here  $\varphi = c\phi$  has dimension mass,
$\sqrt{|\theta^{\mu\nu}|^{-1}}$ is the $SO(5)$-invariant Euclidean volume form on $S^4\subset \R^5$ 
inherited from the symplectic  form $\omega$ on $\C P^3 \sim S^4 \times S^2$, and 
\begin{align}
   \g^{\mu\nu} &= g_{\mu'\nu'}[\theta^{\mu'\mu}\theta^{\nu'\nu}]_{S^2} \ .
   \label{gamma-def}
\end{align}
This is reminiscent of the open string metric 
in the Seiberg-Witten limit \cite{Seiberg:1999vs}.
The crucial volume-form  arises 
because $\Tr\ \sim \int_{\C P^3} \omega^{\wedge 3}\  \sim \int_{S^4} d^4 x\, \sqrt{|\theta^{\mu\nu}|^{-1}}$ is an
integral over the symplectic manifold $\C P^3$. 
Since $|\theta^{\mu\nu}|$ is constant along the internal $S^2$ fiber over $S^4$, the $S^2$
only contributes an irrelevant constant factor which is dropped.
Assuming that low-energy fields $\phi(x)$ are constant along $S^2$, \eq{gamma-def} follows.
 Recasting this kinetic term in the standard covariant metric form,
 we can read off  
 the conformal factor\footnote{The formula given in \cite{Steinacker:2010rh} is modified here due to 
 the averaging over $S^2$.}
\begin{align}
 \ G^{\mu\nu} &= \a\ \g^{\mu\nu}, 
   \qquad \a =  \sqrt{\frac{|\theta^{\mu\nu}|}{|\g^{\mu\nu}|} } \ .
 \label{eff-metric-G}
\end{align}
The average $[\theta^{\mu'\mu}\theta^{\nu'\nu}]_{S^2}$ can be evaluated 
using the  $S^4_N$ formula \cite{Steinacker:2016vgf}
\begin{align}
  \left[\theta^{ab}\theta^{cd}\right]_{S^2} 
  &=\frac{1}{12} \D^4  (P^{ac}_S P^{bd}_S - P^{bc}_S P^{ad}_S + \varepsilon^{abcde} \frac 1{R} x^e) 
\label{average-euclid}
\end{align}
where $\D^2 = 2 r R$ is the space-time uncertainty scale, and
\begin{align}
 P^{ac}_S(x) = \d^{ac} - \frac 1{R^2} x^a x^c, \qquad R^2 = \d_{ab} x^a x^b
\end{align}
is the Euclidean (!) projector $P_S^{ab} \d_{bc} P_S^{cd} = P_S^{ad}$ on the tangent space of $S^4$. 
Note that \eq{average-euclid}
is the unique $SO(5)$- invariant  tensor which reflects the antisymmetry and selfduality\footnote{One must be careful 
not to mix up the Euclidean and Lorentzian aspects. Selfduality of course holds w.r.t. the Euclidean metric.
In the same vein, the trace in \eq{kinetic-term-metric} has nothing to do with the 
target space metric,
and it reduces to the integral over the symplectic $\C P^3 \sim S^4 \times S^2$ as in the Euclidean case.} 
of $\theta^{ab}$.
Then
\begin{align}
 g_{ab} P_S^{ab} &= (-\k^2 +4) - \frac 1{R^2}g_{ac}x^a x^c 
 = (-\k^2 +4) + \frac 1{R^2}(\k^2 x_0^2 - (R^2-x_0^2)) \nn\\
   &= (-\k^2 +3) + (\k^2+1) \cos^2(\eta)
\end{align}
so that 
\begin{align}
\g^{bd} &= \frac{1}{12} \D^4 \Big(g_{ab} P_S^{ab}\, P^{bd}_S
   - g_{ac} P^{bc}_S P^{ad}_S \Big) \nn\\
  &=: \frac{1}{12} \D^4 \tilde\g_{ac} P^{bc}_S P^{ad}_S
\end{align}
with
\begin{align}
 \tilde\g_{ac} &= \big((1+\k^2)\cos^2(\eta) +(3-\k^2) \big) P^S_{ac} - g_{ac} \ .
\end{align}
Before continuing with the evaluation, we consider some special cases.
On the maximal space-like $S^3$ with $x^0 = 0$, the first term vanishes, and
\begin{align}
 \tilde\g_{ab} = -\eta_{ab} = \diag(\k^2,-1-1,-1,-1), \qquad x^0 = 0 \ .
\end{align}
This is indeed Lorentzian, as desired.
In contrast, for  $x_0 = \pm R_x$ we obtain
\begin{align}
 \tilde\g_{ac} &= \frac{1}{12} \D^4 \Big(4 P_{ac}^S - \eta_{ac}\Big) 
  = \frac{1}{12} \D^4 \diag((0,4,4,4,4) - (-\k^2,1,1,1,1))  \nn\\
  &\stackrel{\k^2=3}{=} \frac{1}{12} \D^4 \diag(3,3,3,3,3) 
\end{align}
which is Euclidean. Hence the space-like $\tilde\g_{ii}$ vanish somewhere in between; 
therefore there must be some singularities, which are
tentatively interpreted as Big Bang and Big Crunch. 
In contrast, the  $\tilde\g_{00}$  component never vanishes.

Now we determine the effective metric explicitly. We 
use the local $x^\mu = (\t x^2 x^3 x^4)$ coordinates on $T_p\cM$ at the standard reference point \eq{ref-point},
where $P_S = \diag(1,1,1,1)$. Then
\begin{align}
\tilde\g_{ii} &=  (\k^2+1)\cos^2(\eta) + 2-\k^2  
=  \frac 12\Big((\k^2+1)\cos(2\eta) + (-\k^2+5)\Big)  \nn\\
 &=: 3 c(\eta)  , \qquad i=2,3,4 \nn\\[1ex]
  \tilde\g_{\t\t} &= \big((1+\k^2)\cos^2(\eta) +(3-\k^2) \big) P_{\t\t} - g_{\t\t} \nn\\  
     &= 3
\end{align}
using
\begin{align}
 g_{\t\t} &=   (-\sin\eta,\cos\eta) \diag(-\k^2,1) (-\sin\eta,\cos\eta) \nn\\
  &= (\k^2+1)\cos^2\eta -\k^2 \ .
\end{align}
Therefore 
\begin{align}
 \g^{\mu\nu} &= \frac{1}{4}\D^4 \diag(1,c(\eta),c(\eta),c(\eta))  ,
\end{align}
consistent with the cases $x_0=\pm R_x$ and $x_0=0$ since $c(0) = 1$.

\begin{figure}
\begin{center}
 \includegraphics[width=0.3\textwidth]{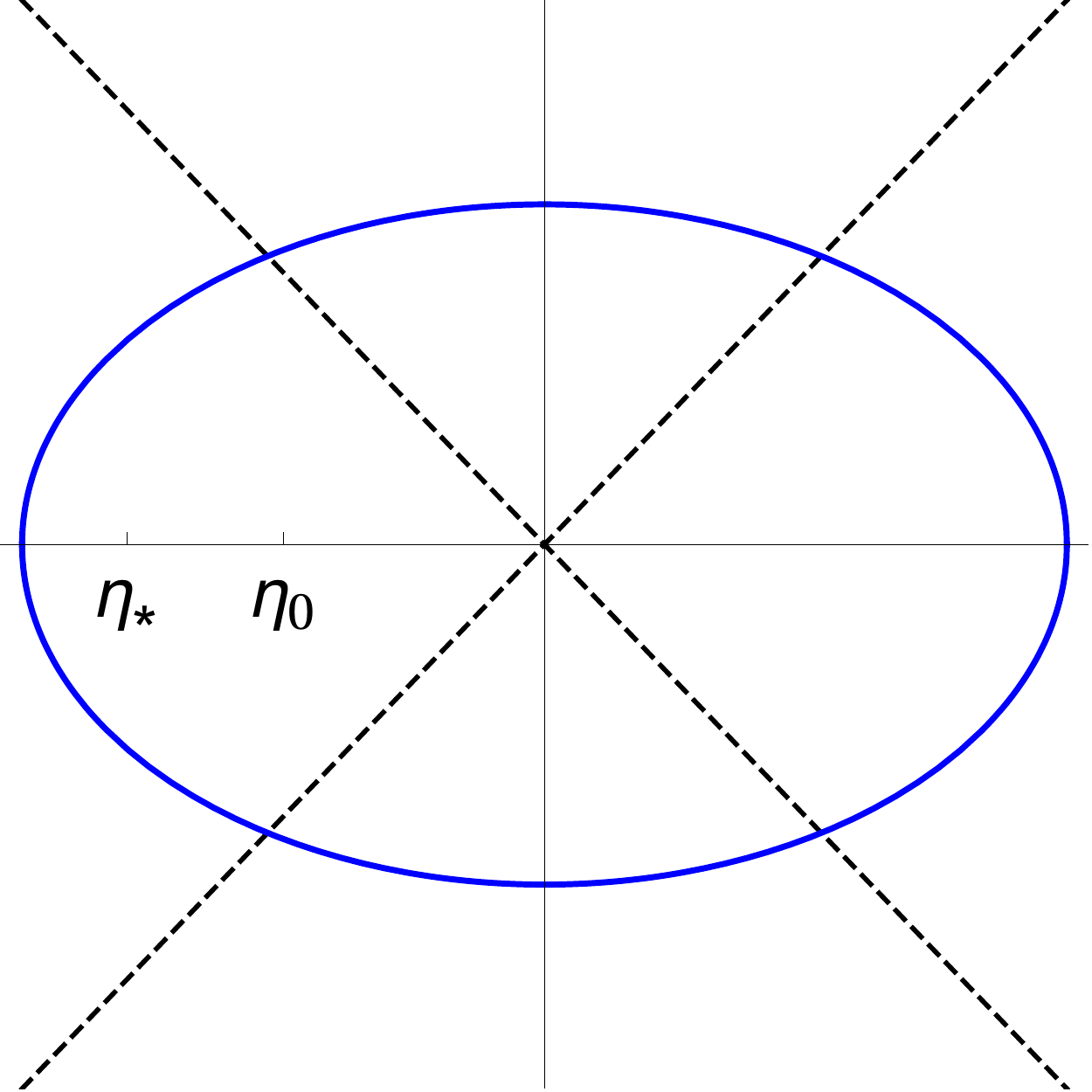}
 \end{center}
 \caption{Schematic picture of the recollapsing universe with  lightcone for $g_{\mu\nu}$, indicating $\eta_*$ and $\eta_0$, .}
 \label{fig:sphere}
\end{figure}

A singularity occurs if the space-components $\tilde\g_{ii}$ change sign, i.e. for $c(\eta_0) = 0$. This happens for 
\begin{align}
 \cos(2\eta_0) = \frac{\k^2-5}{\k^2+1} 
 \label{BB-general-S4}
\end{align}
and is interpreted as Big Bang and Big Crunch. This always has a solution since $\k^2 \geq 3$ \eq{k-bound-S4},
which occurs always {\em after} the signature change \eq{eta-star} for the induced metric,  $\eta_0 >\eta_*$
(in the expanding phase),
as indicated in figure \ref{fig:sphere}.
For $\k^2=3$, this occurs for $\eta_0 = \frac{\pi}{3}$.
Between Big Bang and Big Crunch,  $\g^{\mu\nu}$ has 
signature\footnote{Note that the effective metric has the opposite sign of the induced metric. 
This is due to the Poisson structure which enters $\g^{ab}$.}  $(+---)$ since $c(\eta)<0$.

In the same coordinates, the $SO(5)$ -invariant volume form $\sqrt{|\theta^{\mu\nu}|} \sim \frac{\D^4}{4}$ is constant.
Therefore the conformal factor \eq{eff-metric-G} is
\begin{align}
 \a  =  \sqrt{\frac{|\theta^{\mu\nu}|}{|\g^{\mu\nu}|} }   = \frac{4}{\D^4} |c(\eta)|^{-3/2} \ ,
\end{align}
 and we obtain the effective metric
 \begin{align}
 G^{\mu\nu} &=  |c(\eta)|^{-3/2}\, \diag(1,c(\eta),c(\eta),c(\eta))\nn\\
 G_{\mu\nu} &=  |c(\eta)|^{3/2}\,\diag(1,c(\eta)^{-1},c(\eta)^{-1},c(\eta)^{-1}) \ .
 \label{eff-metric-ellipse}
\end{align}

\paragraph{Scale factor.}

To extract the cosmological evolution, we 
express this metric in FRW coordinates, 
\begin{align}
 ds^2_G = b^2(\eta) d\eta^2 - \tilde a^2(\eta) d\Omega^2
 = d t^2 - a^2(t)d\Omega^2
\end{align}
where $d\Omega^2$ is the length element on a spatial 3-sphere $S^3$ with unit radius.
Thus
\begin{align}
 \tilde a^2(\eta) &= R^2|c(\eta)|^{1/2}\ = a^2(t) , \qquad 
 b^2(\eta) = R^2 |c(\eta)|^{3/2}\, 
\end{align}
in the cosmological  era (with Minkowski signature).
Note that now both $a$ and $b$ vanish at the time $\eta_0$ of the BB, in contrast to the induced metric \eq{g-cartesian}.
We set $R=1$ for simplicity. Then
\begin{align}
 b = a^3
\end{align}
The comoving time parameter $t$ is determined as 
\begin{align}
 \dot\eta &= b^{-1} =  a^{-3}\ .
\end{align}
We can solve this in the cosmological era recalling \eq{BB-general-S4},
\begin{align}
 a^4 = |c(\eta)| 
     &=  \frac 16\big((\k^2 -5) - (\k^2+1)\cos(2\eta)\big),  
\end{align}
which gives
 \begin{align}
 4 a^3 \dot a &= \frac{1}{3}(\k^2+1)\sin(2\eta) \dot\eta 
   = \frac 13(\k^2+1)a^{-3}\sqrt{1-\frac{(6a^4 -\k^2+5)^2}{(\k^2+1)^2}}\, 
 \end{align}
 and finally
  \begin{align}
 \dot a &= \frac 1{12} a^{-6}\sqrt{(\k^2+1)^2 - (6a^4 -\k^2+5)^2} \ .
\end{align}
At early times after the BB, this is approximated by 
\begin{align}
  \dot a  &= c\, a^{-6}, \qquad c = \frac{\sqrt{\k^2-2}}{2\sqrt{3}}\,
\end{align}
which leads to  the initial expansion
\begin{align}
 a(t) \sim  t^{1/7} \ .
 \label{a-initial-t17-S}
\end{align}
Hence the scale parameter $a(t)$ exhibits a very rapid  (but not exponential) initial expansion, 
which slows down naturally.
It reaches a maximum $a_{\rm max}$ at 
\begin{align}
a_{\rm max}^4 &= \frac{\k^2-2}3  \geq \frac 13
\end{align}
after which the universe starts to contract, and eventually collapses in a Big Crunch.
It is decelerating at all times, $\ddot a < 0$.
The Hubble parameter is 
\begin{align}
H(t) = \frac{\dot{a}}{a} &= \frac{1}{12} a^{-7}\sqrt{(\k^2+1)^2 - (6a^4 -\k^2+5)^2} \nn\\
 &\sim \  t^{-1}, \qquad t \approx 0
\end{align}
for the early universe.

\begin{figure}
\begin{center}
 \includegraphics[width=0.5\textwidth]{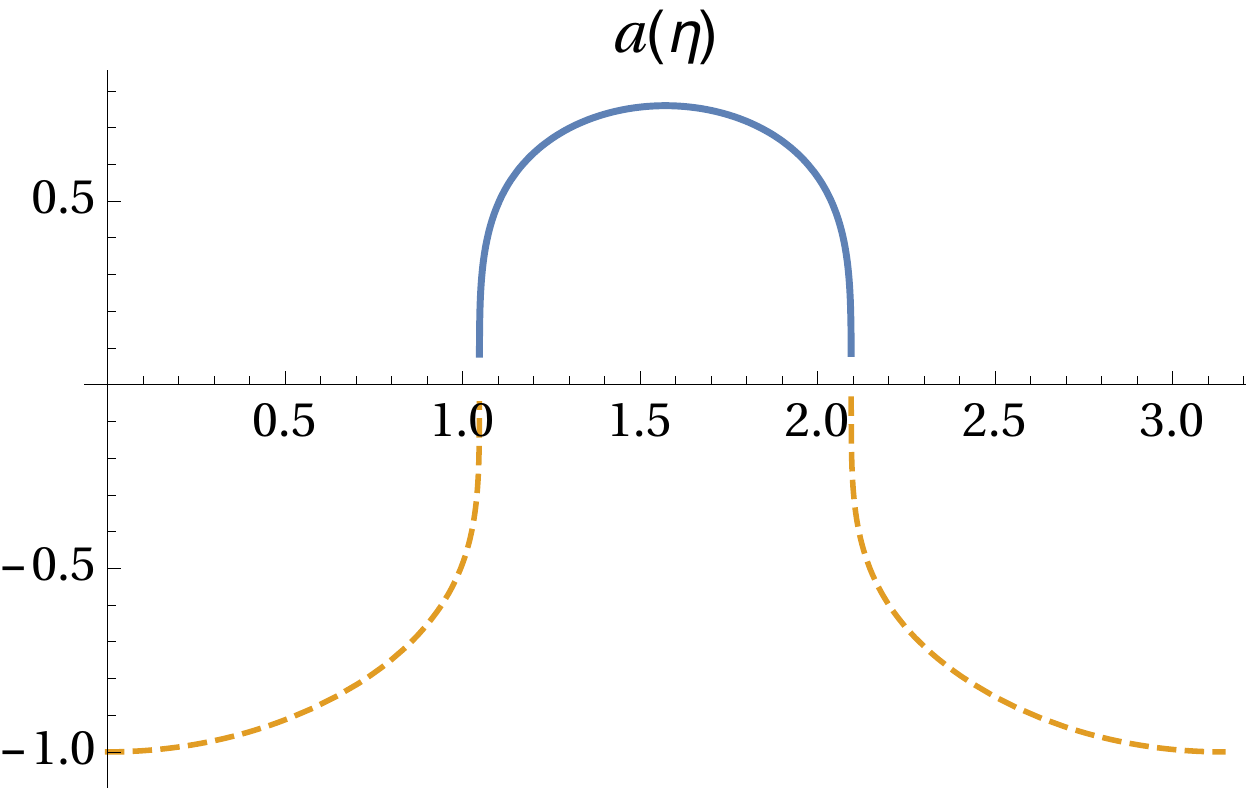}
 \end{center}
 \caption{$a(\eta)$ for the $S^4$ cosmology  with $\k^2 = 3$. 
 The red dashed line describes the Euclidean caps with imaginary $a(\eta)$.}
 \label{fig:a-eta-S4}
\end{figure}

To see what this means from the target space point of view,
we plot $a(\eta)$  in figure \ref{fig:a-eta-S4} as a function of the target space angle $\eta$,
rather than $a(t)$. Then the initial singularity is milder than in the comoving time $t$,
but still manifest.
Note that the scale parameter $a(\eta)$ is imaginary 
for $0\leq\eta< \eta_0$, as indicated in figure \ref{fig:a-eta-S4}.
This gives the Euclidean effective metric for $\eta< \eta_0$.
Since the BB \eq{BB-general-S4} occurs
{\em after} the signature change in the induced metric \eq{eta-star} at $\eta_*$,
there is an era before the BB where the effective metric is Euclidean but the induced metric is Lorentzian.
The induced metric governs the one-loop corrections, which in the IKKT model essentially gives  IIB supergravity
in the 10D bulk, i.e. the closed string sector with short-range  $r^{-8}$ propagators \cite{Ishibashi:1996xs,Chepelev:1997av,Steinacker:2016nsc}.
This should entail {\em some} causal connection  even before the BB (due to the the closed string sector), which 
might  resolve the horizon problem
even in the absence of  standard inflation, and thereby explain the observed uniformity in the CMB.

It is instructive to
compare the effective metric $G_{\mu\nu}$ \eq{eff-metric-ellipse} with the induced metric $g_{\mu\nu}$ \eq{g-cartesian}. The crucial difference
lies in the conformal factor, which is responsible for the expanding BB behavior for $G_{\mu\nu}$ rather than 
just developing a $(0+++)$ degeneracy for $g_{\mu\nu}$. We emphasize again that this conformal factor arises from 
matching the kinetic term in the matrix model action with a covariant metric expression.
Since the matrix model action involves the trace, it incorporates a measure (a density) 
which 
arises from the underlying symplectic manifold $\C P^3$, corresponding to a quantized 4-form flux on $S^4$.

\section{Expanding universe from fuzzy hyperboloids}
\label{sec:open-universe}

Now we repeat the above computation for the  case of a hyperboloid.
We focus on the fuzzy hyperboloid $H^4_n$ as discussed in 
\cite{Hasebe:2012mz,Grosse:2010tm}. 
Analogous to $S^4_N$, this arises from certain irreducible representations 
of the noncompact cousin $SO(1,4)$ of $SO(5)$, and again there are magic representations where this 
structure group is enhanced to $SO(2,4)$.

\subsection{Euclidean fuzzy hyperboloids}

To define fuzzy $H^4_n$, let 
$\eta^{ab} = \diag(-1,1,1,1,1,-1)$ be the invariant metric of $SO(4,2)$, and
$\cM^{ab}$ be hermitian generators of $SO(4,2)$, which satisfy
\begin{align}
  [\cM_{ab},\cM_{cd}] &=i(\eta_{ac}\cM_{bd} - \eta_{ad}\cM_{bc} - \eta_{bc}\cM_{ad} + \eta_{bd}\cM_{ac}) \ .
 \label{M-M-relations-noncompact}
\end{align}
We choose a particular type of (massless discrete series) 
positive-energy unitary irreps\footnote{Strictly speaking there are two versions 
$\cH_{n}^L$ or $\cH_{n}^R$ with opposite ``chirality'', 
but this distinction is irrelevant in the present paper and therefore dropped.} $\cH_{n}$ 
known as ``minireps'' or doubletons \cite{Mack:1975je,Fernando:2009fq}, 
which have the remarkable property that they remain 
irreducible\footnote{This follows from the minimal oscillator construction of $\cH_n$, where all 
$SO(4,2)$ weight multiplicities 
are at most one. Cf. \cite{Mack:1969dg,Mack:1975je,Heidenreich:1980xi}.}
under $SO(4,1) \subset SO(4,2)$. They have positive discrete spectrum 
\begin{align}
 {\rm spec}(\cM^{05}) = \{E_0, E_0+1, ... \}, \qquad E_0 = 1+\frac{n}2 
\end{align}
where the 
eigenspace with lowest eigenvalue of $\cM^{05}$ is an $n + 1$-dimensional irreducible
representation of either $SU (2)_L$ or $SU (2)_R$. Then the hermitian generators
\begin{align}
 X^a &:= r\cM^{a 5}, \qquad a = 0,...,4  \nn\\
   [X^a,X^b] &= i r^2\cM^{ab}  =: i\Theta^{ab} 
\end{align} 
satisfy 
\begin{align} 
 \eta_{ab} X^a X^b &= X^i X^i - X^0 X^0 = - R^2 \one \ 
 \label{hyperboloid-constraint}
\end{align}
with $R^2 = r^2(n^2-4)$ \cite{Hasebe:2012mz}.
Since $X^0 = r \cM^{05} > 0$ has positive spectrum, this describes a one-sided hyperboloid in $\R^{1,4}$,
denoted as $H^4_n$.
Analogous to fuzzy $S^4_N$, the semi-classical geometry underlying  
$H^4_n$ is $\C P^{1,2}$ \cite{Hasebe:2012mz},
which is an $S^2-$ bundle over $H^4$ carrying a canonical symplectic structure. 
In the fuzzy case, this  fiber is  a fuzzy 2-sphere $S^2_n$.
We work again in the semi-classical limit.
It is important to note that the 
induced metric on the hyperboloid $\cM := H^4\subset \R^{1,4}$ is  Euclidean, despite the $SO(4,1)$ isometry. 
This is obvious at the point 
$x=(R,0,0,0,0)$, where the tangent space is $\R^4_{1234}$.

Thus $H^4_n$ has the same local structure as $S^4_n$ in the semi-classical limit, with a
Poisson tensor $\theta^{\mu\nu}(x,\xi)$ transforming as a 2-form under the 
local stabilizer $SO(4)_x$ of any point $x \in \cM$. This
realizes a $S^2$ bundle of self-dual 2-vectors. Then the averaging over $S^2$
 can be achieved using the same local formulas as for $S^4_N$, which will be useful below.

In particular, $H^4_n$ has a finite density of microstates just like $S^4_n$, since
the number of states in $\cH_n$ between two given $X^0$-eigenvalues is finite.
This density can in fact be much smaller than for $S^4_N$, because $n$ is no longer required to be large as we will see.

\subsection{Lorentzian fuzzy hyperboloids}

In analogy to \eq{S4-ansatz}, Lorentzian spaces of the above type with suitable 
rescaling of the generators $X^a$ are solutions of the same matrix model \eq{bosonic-action},
for suitable mass parameters.
Thus we look for solutions  of the mass-deformed matrix model
given by rescaled generators $Y^a, \ a\in\{0,...,4\}$
\begin{align}
   Y^i &=  X^i , \quad \mbox{for} \ \  i = 1,...,4, \qquad  Y^0 = \k X^5 \ .
 \label{H4-ansatz}
\end{align}
The transform as vectors of $SO(4)$.
We can again rewrite the model in the $X^a$ coordinates as in \eq{bosonic-action-X}, with
\begin{align}
g_{ab} = \diag(-\k^2,1,1,1,1).
\label{etatilde-lorentz}
\end{align}
 The commutation relations now give
\begin{align}
  [X^a,[X^a,X^b]] &= i r^2 [X^a,\cM^{ab}] = i r^2 [\cM^{ba},X^a] \qquad\mbox{(no sum)}  \nn\\
    &=  r^2 \left\{\begin{array}{ll}
                X^b, & b\neq a \neq 0 \\
               -  X^b, & b\neq a = 0 \\
               0, & b = a
              \end{array}\right. \ .
\end{align}
Thus the equations of motion
\begin{align}
 (\Box_X + \tilde m_0^2) X^0 = 0  = (\Box_X +m^2) X^i 
\end{align}
 reduce to
\begin{align}
 4 r^2 + \tilde m_0^2 = 0 = (3+\k^2)r^2  + m^2 \ .
\end{align}
Now both mass terms need to be  negative
$m_0^2 < 0$ and $m^2 < 0$, with
\begin{align}
 \frac{3+\k^2}4 = \frac{m^2}{\tilde m_0^2} =  \frac{m^2}{\k^2 m_0^2} \ .
\end{align}
We will see that 
a Lorentzian effective metric arises for $\k^2 > 1$ \eq{eta-0-hyperbel}, i.e.
for $m^2 < m_0^2 < 0$. Then
 $H^4_n$ with $X^a X^b \eta_{ab} = - R^2$ \eq{hyperboloid-constraint} is 
indeed a solution of the  matrix model with Lorentzian target space metric   
$g_{ab}$ \eq{eta-tilde}, and $Y^a$ is a solution of \eq{bosonic-action}.

One may worry about possible instabilities 
in the presence of negative masses. 
However, these mass terms are of cosmological scale, and therefore extremely small.
Moreover as shown in the case of $S^4_N$ \cite{Steinacker:2015dra}, even a positive bare mass term may lead to 
a  radius stabilization  at one loop, as the quantum effective action mimics
a negative mass for the radial parameter(s). Thus one may hope that quantum effects stabilize the present solution
even in the presence of positive but 
different masses. The computation of the effective metric below
would then essentially go through.

\paragraph{Induced metric.}

Again 
we  compute first the induced metric $g_{\mu\nu}$ on $H^4\subset \R^{1,4}$, which clearly has Euclidean signature 
at $x_0 = R$. Using the  space-like $SO(4)$ symmetry,
we can choose a standard reference point 
\begin{align}
 x=(x_0,x_1,0,0,0) = R(\cosh(\eta),\sinh(\eta),0,0,0) \ \in H^4 \subset \R^{4,1} , \qquad x_0^2 - x_1^2 = R^2 \ .
 \label{ref-point-H}
\end{align}
and use the tangential coordinate 
\begin{align}
 \tau = R \eta
\end{align}
which points in the $x^0 x^1$ direction,  
\begin{align}
 \frac{d}{d\t} x = R(\sinh(\eta),\cosh(\eta),0,0,0) \ . 
\end{align}
Then the induced metric is 
\begin{align}
 g_{\mu\nu} = \diag(\cosh^2(\eta) - \k^2\sinh^2(\eta),1,1,1)
 \label{g-cartesian-H}
\end{align}
in local 
$x^\mu = (\t x^2 x^3 x^4)$ coordinates on $T_p\cM$ for the standard reference point \eq{ref-point},
or
\begin{align}
 ds^2_g &= R^2 (\cosh^2(\eta) - \k^2\sinh^2(\eta))d\eta^2 + R^2\sinh^2(\eta) d\Omega_3  \nn\\
  &= \b^2(\eta) d\eta^2 + \a^2(\eta)d\Omega_3^2 
\end{align}
in FRW coordinates 
with
\begin{align}
 \b^2(\eta) =  \frac 12 R^2 \big((-\k^2+1)\cosh(2\eta) +\k^2+1\big), 
  \qquad \a^2 = \frac 12 R^2(1-\cosh(2\eta)).
\end{align}
Clearly $\a\geq 0$ vanishes only for $x^0 = R$ where $\eta=0$. In contrast, 
$\b(\eta_*) = 0$ vanishes if
\begin{align}
 \cosh(2\eta_*) = \frac{\k^2+1}{\k^2-1} \ .
 \label{eta-star-H}
\end{align}
So again there is an interesting transition from Euclidean to Minkowski signature, but
the associated singularity cannot be interpreted as Big Bang. Rather,
the Big Bang will arise for the effective metric.

\paragraph{Effective metric.}

\begin{figure}
\begin{center}
 \includegraphics[width=0.3\textwidth]{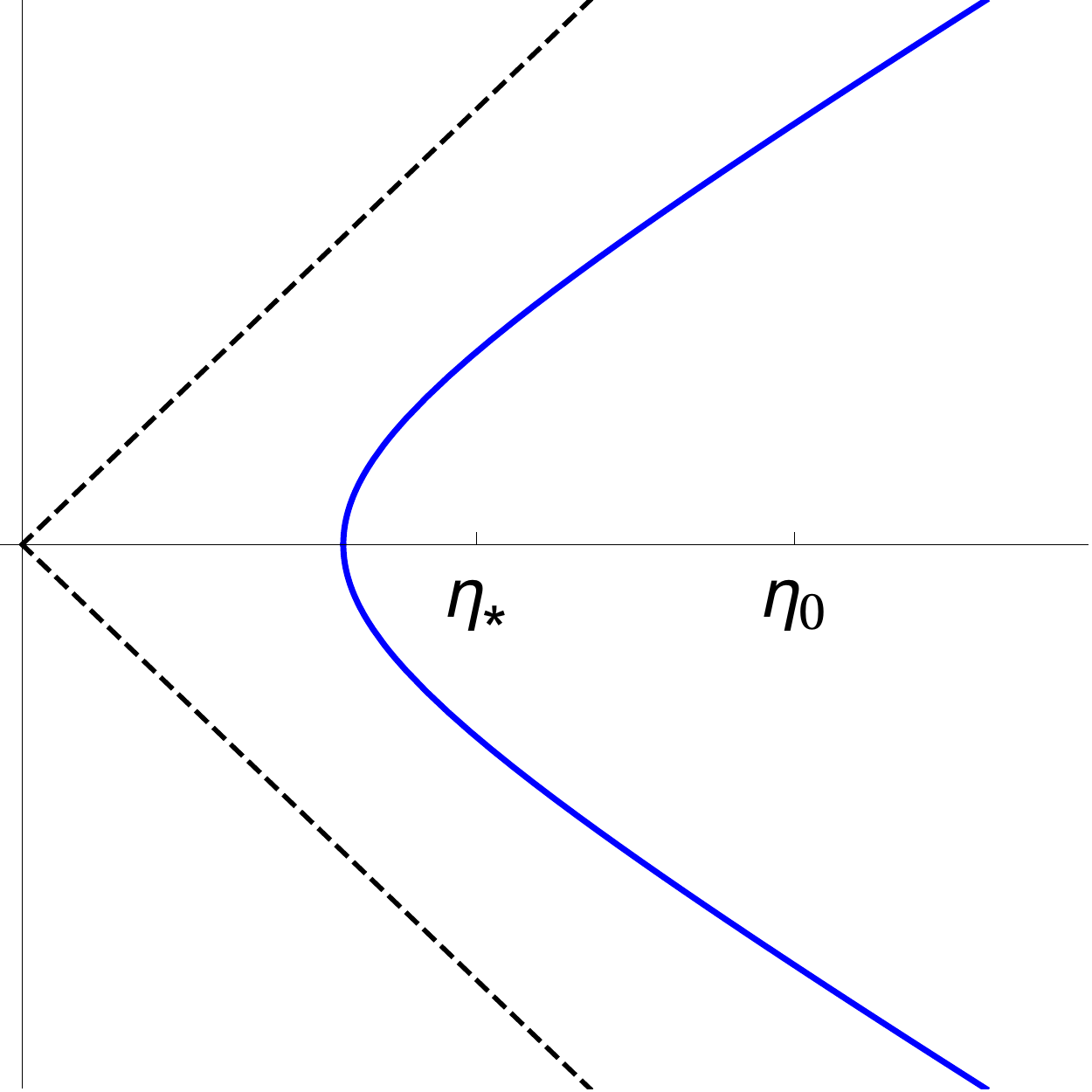}
 \end{center}
 \caption{Schematic picture of the expanding universe  with lightcone for $g_{\mu\nu}$, indicating $\eta_*$ and $\eta_0$.}
 \label{fig:hyperbola}
\end{figure}

To obtain the effective metric \eq{eff-metric-G} on $H^4$, we need to compute the average 
$\left[\theta^{ab} \theta^{cd}\right]_{S^2}$ for $H^4$.
Recall that in the semi-classical limit, the $X^a \sim x^a$ provide an embedding of $H^4$ as a Euclidean 
hyperboloid in $\R^{1,4}$,
with (anti)selfdual  $\theta^{\mu\nu}$  describing an internal $S^2$ fiber.
Therefore the averaging over this fiber is achieved  as before\footnote{This is the reason 
for using the $X^a$ coordinates. The Minkowskian metric $\eta^{ab}$ plays no role for this averaging.}
via
\begin{align}
  \left[\theta^{ab} \theta^{cd}\right]_{S^2} 
  &=\frac{1}{12} \D^4  (P^{ac}_H P^{bd}_H - P^{bc}_H P^{ad}_H \pm \varepsilon^{abcde} \frac 1{R} x^e) ,
\label{average-Mink}
\end{align}
where now
\begin{align}
 P^{ac}_H(x) = \eta^{ac} + \frac 1{R^2} x^a x^c, \qquad \eta_{ab} x^a x^b =  -R^2 
\end{align}
is the $SO(4,1)$-invariant Euclidean projector 
on the tangent space of $H^4$ (which is Euclidean w.r.t. $\eta_{ab}$). 
Note that
\eq{average-Mink} is the unique $SO(4,1)$- invariant  tensor which reflects the antisymmetry and 
(anti)selfduality of $\theta^{ab}$.
Again we shall evaluate this at the reference point \eq{ref-point-H}  on $H^4$.
Then
\begin{align}
g_{ab} P_H^{ab} &= g_{ab} \eta^{ab} + \frac{1}{R^2}g_{ab}x^a x^b \nn\\
&= (\k^2 +4) + \frac{1}{R^2}(-\k^2 x_0^2 + x_0^2 - R^2)   \nn\\
&= (\k^2 +3) -(\k^2-1) \cosh^2(\eta)
\end{align}
so that 
\begin{align}
\g^{bd} &= \frac{1}{12} \D^4 \Big((g_{ab} P_H^{ab}) P^{bd}_H
   - g_{ac} P^{bc}_H P^{ad}_H \Big) \nn\\
  &=: \frac{1}{12} \D^4 \tilde\g_{ac} P^{bc}_H P^{ad}_H 
\end{align}
where 
\begin{align}
 \tilde\g_{ac} = ((\k^2 +3) -(\k^2-1) \cosh^2(\eta)) P_{ac}^H - g_{ac} .
\end{align}
For the ``undeformed'' case $\k^2=1$, we recover $\tilde\g_{ac} = 4 P_{ac}^H - g_{ac} = 3  P_{ac}^H$.
For $\eta=  0$ i.e. $(x_0 = R, \ x_1=0)$ this is Euclidean,
\begin{align}
 \tilde\g_{ac} = 4 P_{ac}^H - g_{ac} = \diag(\k^2,3,3,3,3) .
\end{align}
More generally, we compute in the local $(\t x^2 x^3 x^4)$ coordinates
\begin{align}
\tilde\g_{ii} &=  \k^2 +2 -(\k^2-1) \cosh^2(\eta)   
 =  \frac 12\big(\k^2 +5 - (\k^2-1)\cosh(2\eta)\big)  \nn\\ 
 &=: 3 c(\eta)   ,   \qquad i=2,3,4  \nn\\[1ex]
\tilde\g_{\t\t} &=  ((\k^2 +3) -(\k^2-1) \cosh^2(\eta)) P_{\t\t}^H - g_{\t\t}  \nn\\
   &= 3 
\end{align}
since $P_{\t\t}^H=1$ and 
\begin{align}
 g_{\t\t} &= (\sinh(\eta),\cosh(\eta)) \diag(-\k^2,1) (\sinh(\eta),\cosh(\eta)) \nn\\
   &= \k^2 + (-\k^2+1)\cosh^2\eta \ .
\end{align}
Therefore $\del_\tau$ is always space-like, and
 \begin{align}
 \g^{\mu\nu} = \frac 1{4} \D^4 (1,c(\eta) ,c(\eta),c(\eta)) 
\end{align}
in the above coordinates. This is
consistent with $x^i=0$ since $c(0) = 1$.
The effective metric changes signature at $c(\eta_0) = 0$ i.e.
\begin{align}
\cosh(2\eta_0) =
\frac{\k^2 +5}{\k^2-1} 
\label{eta-0-hyperbel}
\end{align}
provided $\k^2>1$,
and the metric is Lorentzian for $\eta > \eta_0$ with $c(\eta)<0$.
Again, this happens {\em after} the signature change for the induced metric \eq{eta-star-H},
as indicated in figure \ref{fig:hyperbola}.
From a target space point of view,
the size of the universe at the BB may or may not be large depending on $\k$, but it is 
small in the effective metric.

In the same coordinates, the $SO(1,4)$ -invariant volume form $\sqrt{|\theta^{\mu\nu}|} \sim \frac{\D^4}{4}$ is constant.
Hence the conformal factor \eq{eff-metric-G} is
\begin{align}
 \a  =  \sqrt{\frac{|\theta^{\mu\nu}|}{|\g^{\mu\nu}|} }   = \frac{4}{\D^4} |c(\eta)|^{-3/2}
\end{align}
 and the effective metric is obtained as
 \begin{align}
 G_{\mu\nu} &=  |c(\eta)|^{3/2}\,\diag(1,c(\eta)^{-1},c(\eta)^{-1},c(\eta)^{-1}) \ .
 \label{eff-metric-hyperbel}
\end{align}
\paragraph{Scale factor.}

Extracting the cosmological evolution proceeds as in section \ref{sec:lorentzian-S4}.
We express the metric in FRW coordinates, 
\begin{align}
 ds^2_G = b^2(\eta) d\eta^2 - \tilde a^2(\eta) d\Omega^2
 = d t^2 - a^2(t)d\Omega^2
\end{align}
where $d\Omega^2$ is the length element on a spatial 3-sphere $S^3$ with unit radius.
Thus
\begin{align}
 \tilde a^2(\eta) &= R^2|c(\eta)|^{1/2}\ = a^2(t) , \qquad 
 b^2(\eta) = R^2 |c(\eta)|^{3/2}\, 
\end{align}
in the cosmological  era (with Minkowski signature).
Again both $a$ and $b$ vanish at the time $\eta_0$ of the BB, in contrast to the induced metric.
We set $R=1$ for simplicity. Then
\begin{align}
 b = a^3 \ .
\end{align}
The comoving time parameter $t$ is determined from 
\begin{align}
 \dot\eta &= b^{-1} =  a^{-3}\ .
\end{align}
We can solve this again in the cosmological era
\begin{align}
 a^4 &= |c(\eta)| =  \frac 16\big(-(\k^2 +5) + (\k^2-1)\cosh(2\eta)\big) \ , 
\end{align}
which gives
 \begin{align}
 \dot a &= \frac 1{12} a^{-6}\sqrt{(\k^2+5+6a^4)^2-(\k^2-1)^2} \ .
\end{align}
This shows  the same initial  $a\sim t^{1/7}$ expansion as in \eq{a-initial-t17-S}.
However for large $t$, we obtain
\begin{align}
  \dot a 
   \ \approx \ \frac 1{2} a^{-2}
   \label{adot-late}
\end{align}
so that 
\begin{align}
 a(t) \ \sim \   t ^{1/3}
\end{align}
for the late-time evolution. The expansion is somewhat slower than for a matter-dominated universe,
which would be $a(t) \sim t^{2/3}$.
The Hubble parameter for large $t$ is 
\begin{align}
 H = \frac{\dot a}{a} \approx \frac{1}{2} a^{-3} \sim t ^{-1} \ ,
 \label{H-late}
\end{align}
and it is again decelerating at all times, $\ddot a < 0$.

\begin{figure}
\begin{center}
 \includegraphics[width=0.4\textwidth]{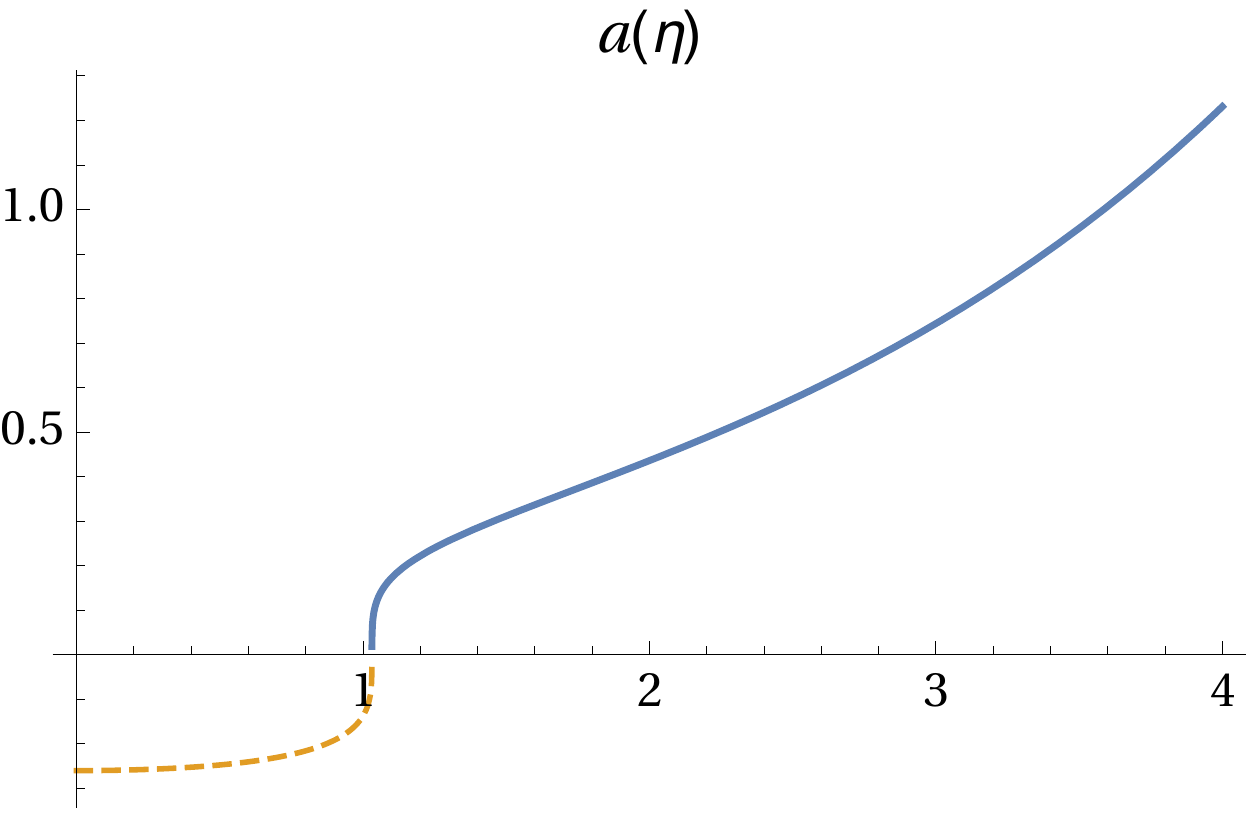} \hspace{1cm}
 \includegraphics[width=0.4\textwidth]{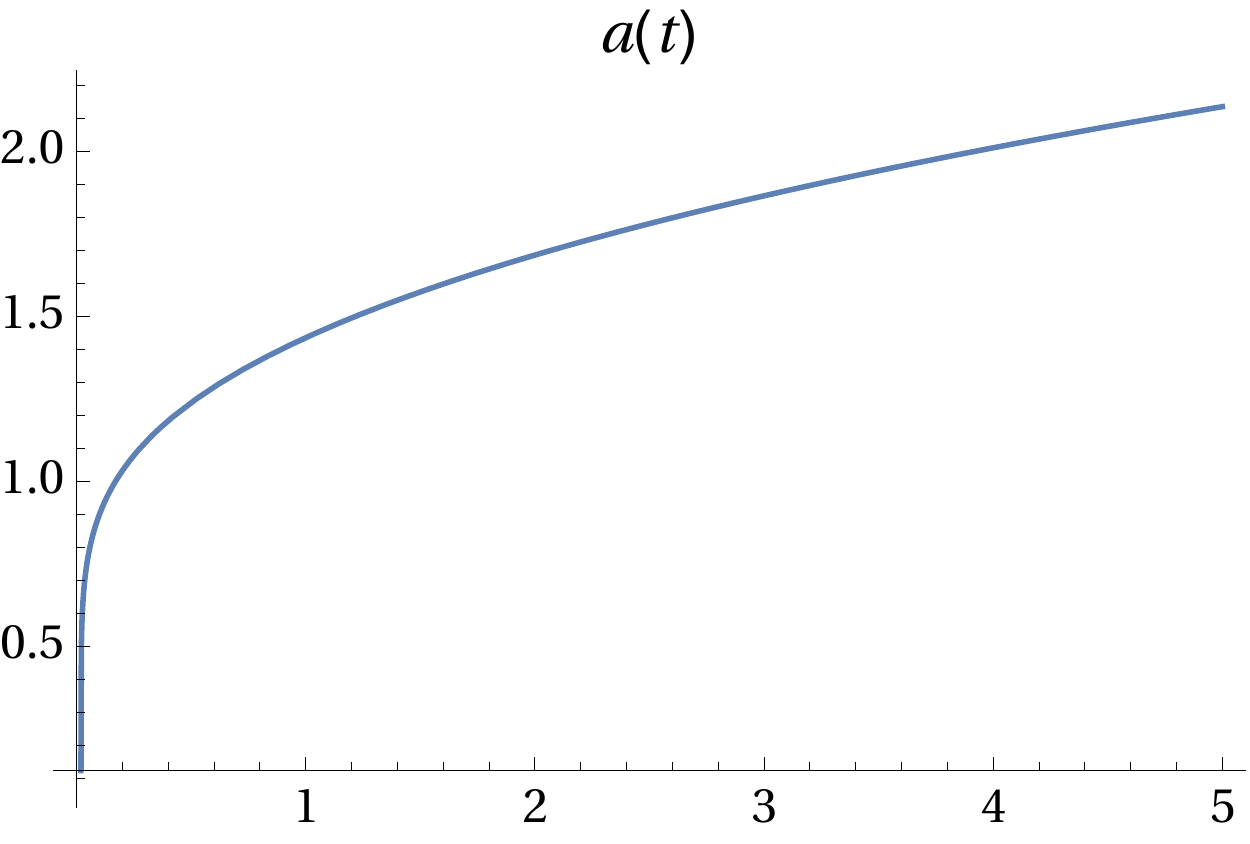}
 \end{center}
 \caption{$a(\eta)$ and $a(t)$ for the $H^4$ cosmology with $\k^2 = 3$. The red dashed line is the Euclidean 
 era with imaginary $a(\eta)$.}
 \label{fig:a-eta-H4}
\end{figure}

To illustrate the expansion,
we plot $a(\eta)$ as well as $a(t)$ in figure \ref{fig:a-eta-H4}. 
From the target space  point of view,  
the BB singularity of $a(\eta)$ is milder than in the $t$ variables,
but still manifest.
Again, the scale parameter $a(\eta)$ is imaginary before the BB 
for $0\leq\eta< \eta_0$, covering the entire $H^4$.
Accordingly, the effective metric is Euclidean for $\eta< \eta_0$.
The apparent acceleration of $a(\eta)$ in figure  \ref{fig:a-eta-H4} is however an artifact,
and there is no acceleration in the comoving time $a(t)$.
Nevertheless, it suggests that  some mild corrections of the metric may easily modify this conclusion;
see e.g. \cite{Chaney:2015ktw} for a related discussion in 2 dimensions.
Some other aspects of this solutions will be discussed below.

\subsection{Outlook}
\label{sec:outlook}

\paragraph{Excitation modes.}


The above solutions define not only geometrical space-times $\cM$;  the bosonic and fermionic 
excitation modes on these backgrounds in the matrix model define gauge theories living on $\cM$.
These fluctuations can be understood in terms of  the noncommutative algebra of functions, 
which is $End(\cH) \sim Fun(\C P^{1,2})$ for $H^4_n$, and 
$End(\cH) \sim Fun(\C P^3)$ for $S^4_N$. Since these are  
equivariant (``twisted'') bundles over  $\cM^4$, 
the harmonics on the fiber $S^2$ lead to higher spin modes on $\cM$
(in contrast to  Kaluza-Klein modes which arise on ordinary compactifications).
Explicitly,  functions $\Phi \in End(\cH_n)$ can be expanded in the form
\begin{align}
 \Phi = \phi(x) + \phi_{ab}(x) \cM^{ab} + ...  \ .
 \label{phi-modes}
\end{align}
This amounts to a decomposition into higher spin modes as in \cite{Sperling:2017gmy,Steinacker:2016vgf},
whose propagation is governed by $\Box_Y$ \eq{Box-Y}, hence by the effective metric $G_{\mu\nu}$.
In particular, 
the tangential fluctuations $Y^\mu + \cA^\mu$ include the  modes
\begin{align*}
 \cA^\mu = \theta^{\mu\nu} h_{\nu\rho}(x) P^\rho
\end{align*}
where $P^\mu \in \mso(4,1)$ is the local generator of translations, cf. \cite{Steinacker:2016vgf}.
The symmetric part of $h_{\mu\nu}$ could naturally play the role of the spin 2 graviton. 
Whether or not this leads to an acceptable gravity will be examined elsewhere.

\paragraph{De Sitter and other solutions.}

It is natural to wonder about de Sitter branes. 
There are indeed candidates for fuzzy de Sitter space based on the principal series representations 
of $SO(4,1)$  \cite{Buric:2017yes,Gazeau:2006hj,Hasebe:2012mz,Jurman:2013ota}, 
some of which should be solutions of the matrix model for $m_0^2 = m^2$; see also
\cite{Heckman:2014xha} for related work. 
However then the effective metric would coincide with the induced one, without  BB. 
Even for $m_0^2 \neq m^2$, it is hard to see how a BB might arise in this case.
Therefore we will not consider this case in the present paper.
There are also mathematical issues, such as the expected non-compact nature of the  
internal space, which makes the averaging procedure problematic for de Sitter-type spaces.

It may also be possible to find a twisted embedding along the lines
of \cite{Klammer:2009ku} to obtain a matrix realization of a covariant space-time with big bounce\footnote{Note 
that the mechanism in \cite{Klammer:2009ku} is different from the present one and essentially 
relies on the embedding metric, which can be the effective one only assuming a certain complexification.}.

\section{Discussion}
\label{sec:discussion}

We presented a novel and simple mechanism how a cosmological Big Bang could arise in the context 
of Yang-Mills matrix models. The BB arises from a signature change in the effective metric 
on  noncommutative space-time branes embedded in Lorentzian target space, taking into account the quantized 4-volume form.
The underlying brane is completely regular, at least in the present simplified  treatment. 
The rapid initial expansion arises from a singular conformal factor, which follows from
the quantized flux in conjunction with the signature change.
There is a period ``before'' the BB where the effective metric is Euclidean but the embedding 
(``closed string``) metric still has Minkowski signature.
One may  hope that this  helps to avoid the horizon problem,
possibly even in the absence of exponential inflation.
The initial sector of the brane is a Euclidean cap. This is somewhat reminiscent 
of an instanton\footnote{This aspect is somewhat reminiscent of Vilenkin's ``tunneling from nothing'' 
proposal \cite{Vilenkin:1982de}.  I would like to thank H. Kawai for pointing this out.}, 
however the path integral \eq{path-integral} is always over $e^{i S}$.

Note that the mechanism does not apply to more traditional brane-world scenarios,
where the effective metric is the induced (pull-back) metric from the target space metric. In such a scenario, 
a signature change would not lead to a rapid initial expansion.

The late-time behavior found in the solutions under consideration is different from the currently accepted $\L$CDM model,
even for the solution based on $H^4$ which is expanding forever.
For example, we can compute the age of universe in terms of the 
present Hubble parameter, assuming that the early phase of the universe is negligible:
\begin{align}
 t = \int\frac{da}{a H(a)} 
  = \frac{1}{3 H(t)} \ .
\end{align}
This deviates from the accepted values by a factor $\approx 3$. Also, 
 equation \eq{H-late} is rather strange compared with the Friedmann equations.
 On the other hand,
the analysis of the model is very crude:
the influence of matter or radiation, and even gravity in the ordinary sense, are completely ignored.
It is therefore  remarkable that semi-realistic cosmologies  including a BB nevertheless arise,
without any reference to the Friedmann equations and GR. 
Hence basic cosmology might have a simple and robust origin in this scenario.

Of course a space-time by itself does not provide a full cosmology.
To obtain interesting physics, gravity must of course be present, at least for 
intermediate scales. As explained in section \ref{sec:outlook}, spin 2 modes 
which could  play the role of gravitons do indeed arise, however this needs to be re-examined carefully
for the present Lorentzian backgrounds\footnote{While for undeformed $S^4_N$ this does not appear to be realistic
\cite{Sperling:2017gmy}, the case of deformed $H^4_n$ with small $n$ looks very promising.}.
Along with the other excitation modes,
this will clearly affect the expansion of the universe.
Loop corrections will also modify the geometry of the brane solution,
e.g. via corrections to the mass parameters in the model, which might even depend on time.
Fuzzy extra dimensions realized along the lines of \cite{Sperling:2017dts} may also affect the expansion.
Finally, the BB entails high temperatures, which will certainly have an impact on the early expansion.
Therefore the quantitative results should  be taken with much caution.

There is also a more basic issue which needs to be clarified. We have determined the 
conformal factor of the metric by matching the kinetic term with the standard covariant metric form \eq{kinetic-term-metric}.
If we would repeat this procedure for a naive mass term, we would obtain a different conformal factor.
This may be reconciled noting that the mass terms for matter should in fact {\em not} arise from the bare
matrix model but from spontaneous symmetry breaking as in the standard model, which would presumably lead 
to a consistent picture. Therefore the present approach seems justified; 
note also that there is no issue for gauge fields due to conformal invariance. 
Nevertheless, the treatment of the conformal factor may need some refinement, which could have a non-trivial effect on the 
late-time cosmology.

From a more formal perspective, the present solutions are also very interesting. In particular, the 
 solutions provide simple examples for a homogeneous and isotropic quantum space-times with Minkowski signature,
with intrinsic IR and  UV cutoff\footnote{There is some superficial similarity with \cite{Chamseddine:2014nxa}, 
however that approach is still based on 
the infinite-dimensional algebra of functions on a classical Euclidean manifold.} in the $S^4_N$ case, and
a UV cutoff in the $H^4_n$ case.
Hence the  mathematical tools and techniques for field theory on such a space can  be worked out,
including the appropriate boundary conditions and the $i\varepsilon$ prescription for loop integrals.
This would allow to compute quantum corrections to the geometry as well as for field theory
in a clear-cut way, notably for the supersymmetric IKKT model.
In particular, the stabilization mechanism in \cite{Steinacker:2015dra} should apply in some way.
On the other hand,
the $H^4_n$ solution is perhaps the most reasonable noncommutative cosmological solution available up to now, and 
it is very promising from the point of view of emergent gravity, due to the presence of spin 2 modes.

In summary,
the main message of this paper is a conceptually very appealing mechanism for a Big Bang within the matrix model,
based on a quantum structure of  space-time.
A large universe arises  quite naturally,  determined only by a discrete choice of the representation, 
as well as a parameter $\k$ which can be of order 1.
The  mechanism is robust and essentially classical (unlike e.g. Vilenkin's 
``universe from nothing'' proposal \cite{Vilenkin:1982de} for the BB), 
and does not rely on general relativity. 
However, a more detailed understanding of the associated physics is required, which needs much more work.

%

%

\paragraph{Acknowledgements.}

I would like to thank in particular Hikaru Kawai and Kentaroh Yoshida for very useful discussions, hospitality 
and support during a stay at Kyoto University, where part of this work was done.
I would also like to thank J. Nishimura and M. Sperling for useful discussions. 
This work was made possible by the Austrian Science Fund (FWF) grant P28590. 
The Action MP1405 QSPACE from the European Cooperation in Science and Technology (COST) also
provided support in the context of this work.

\appendix

\section{No finite-dimensional solutions of massless Lorentzian Yang-Mills matrix models}
\label{sec:mass-lemma}

We show the following simple result, which explains the presence of $m^2$ in \eq{eom-lorentzian-M}:

{\bf Lemma:}

Let $\Box_X = \eta_{ab}[X^a,[X^b,.]]$ for $\eta_{ab}=\diag(-1,1,...,1)$.
Assume that $X^0,X^i$ are finite-dimensional hermitian matrices which are solution of 
\begin{align}
 \Box_X X^0 = 0 = \Box_X X^i \ .
\end{align}
Then all matrices $X_i,X_0$ commute with each other.

{\bf Proof:} 

The eom for $X^0$ implies
\begin{align}
- tr (X^0 \Box_X X^0) = 0 = \sum_i tr(([X^0,X^i])^2 = \sum_i tr A_i^2
\end{align}
where $A_i := i[X_0,X_i] = A_i^\dagger$ is hermitian. It follows that 
\begin{align}
 0 = tr A_i^2 \qquad \forall i
\end{align}
and therefore $A_i=0$. This means that 
\begin{align}
 [X_0,X_i] = 0 \ .
\end{align}
Then the equations of motion for $X_i$ imply 
\begin{align}
 \pm tr X^i \Box_X X^i = 0 = \sum_j tr(([X^j,X^i])^2  \qquad \mbox{for each}\ i \ .
\end{align}
This implies
\begin{align}
 [X_i,X_j] = 0  \qquad \forall i,j
\end{align}
as before, and all matrices commute.

However, there are finite-dimensional non-commutative solutions in the presence of masses, 
as illustrated in this paper.

\stoptocwriting

\resumetocwriting


\begin{thebibliography}{99}


\bibitem{Ishibashi:1996xs} 
  N.~Ishibashi, H.~Kawai, Y.~Kitazawa and A.~Tsuchiya,
  ``A Large N reduced model as superstring,''
  Nucl.\ Phys.\ B {\bf 498}, 467 (1997)
  [hep-th/9612115]. 

  
\bibitem{Chepelev:1997av} 
  I.~Chepelev and A.~A.~Tseytlin,
  ``Interactions of type IIB D-branes from D instanton matrix model,''
  Nucl.\ Phys.\ B {\bf 511}, 629 (1998)
  [hep-th/9705120];
 M.~R.~Douglas and W.~Taylor,
  ``Branes in the bulk of Anti-de Sitter space,''
  hep-th/9807225.
  
  \bibitem{Steinacker:2016nsc} 
  H.~C.~Steinacker,
  ``String states, loops and effective actions in noncommutative field theory and matrix models,''
  Nucl.\ Phys.\ B {\bf 910}, 346 (2016)
  [arXiv:1606.00646 [hep-th]].
  
  
  
\bibitem{Aoki:1999vr} 
  H.~Aoki, N.~Ishibashi, S.~Iso, H.~Kawai, Y.~Kitazawa and T.~Tada,
  ``Noncommutative Yang-Mills in IIB matrix model,''
  Nucl.\ Phys.\ B {\bf 565}, 176 (2000)
  [hep-th/9908141].
  
  
  
\bibitem{Szabo:2001kg} 
  R.~J.~Szabo,
  ``Quantum field theory on noncommutative spaces,''
  Phys.\ Rept.\  {\bf 378}, 207 (2003)
  [hep-th/0109162].
  
  
  
\bibitem{Minwalla:1999px} 
  S.~Minwalla, M.~Van Raamsdonk and N.~Seiberg,
  ``Noncommutative perturbative dynamics,''
  JHEP {\bf 0002}, 020 (2000)
  [hep-th/9912072];
  Y.~Kinar, G.~Lifschytz and J.~Sonnenschein,
  ``UV / IR connection: A Matrix perspective,''
  JHEP {\bf 0108}, 001 (2001)
  [hep-th/0105089].
  
  
  
  
\bibitem{Steinacker:2010rh} 
  H.~Steinacker,
  ``Emergent Geometry and Gravity from Matrix Models: an Introduction,''
  Class.\ Quant.\ Grav.\  {\bf 27}, 133001 (2010)
  
  
\bibitem{Steinacker:2016vgf} 
  H.~C.~Steinacker,
  ``Emergent gravity on covariant quantum spaces in the IKKT model,''
  JHEP {\bf 1612}, 156 (2016)
  [arXiv:1606.00769 [hep-th]]
  
  
\bibitem{Seiberg:1999vs} 
  N.~Seiberg and E.~Witten,
  ``String theory and noncommutative geometry,''
  JHEP {\bf 9909}, 032 (1999)
  doi:10.1088/1126-6708/1999/09/032
  [hep-th/9908142].
  
 \bibitem{Jurman:2013ota} 
  D.~Jurman and H.~Steinacker,
  ``2D fuzzy Anti-de Sitter space from matrix models,''
  JHEP {\bf 1401}, 100 (2014)
  [arXiv:1309.1598 [hep-th]].
  
  
 \bibitem{Chaney:2015ktw}  
 A.~Chaney, L.~Lu and A.~Stern,
  ``Matrix Model Approach to Cosmology,''
  Phys.\ Rev.\ D {\bf 93}, no. 6, 064074 (2016)
  [arXiv:1511.06816 [hep-th]];
  
\bibitem{Steinacker:2011wb} 
  H.~Steinacker,
  ``Split noncommutativity and compactified brane solutions in matrix models,''
  Prog.\ Theor.\ Phys.\  {\bf 126}, 613 (2011)
  [arXiv:1106.6153 [hep-th]].
  
 
\bibitem{Kim:2012mw}  
  S.~W.~Kim, J.~Nishimura and A.~Tsuchiya,
  ``Late time behaviors of the expanding universe in the IIB matrix model,''
  JHEP {\bf 1210}, 147 (2012)
  [arXiv:1208.0711 [hep-th]];
  S.~W.~Kim, J.~Nishimura and A.~Tsuchiya,
  ``Expanding (3+1)-dimensional universe from a Lorentzian matrix model for superstring theory in (9+1)-dimensions,''
  Phys.\ Rev.\ Lett.\  {\bf 108} (2012) 011601
  [arXiv:1108.1540 [hep-th]];
  S.~W.~Kim, J.~Nishimura and A.~Tsuchiya,
  ``Expanding universe as a classical solution in the Lorentzian matrix model for nonperturbative superstring theory,''
  Phys.\ Rev.\ D {\bf 86}, 027901 (2012)
  [arXiv:1110.4803 [hep-th]].

  
\bibitem{Chaney:2015mfa} 
  A.~Chaney, L.~Lu and A.~Stern,
  ``Lorentzian Fuzzy Spheres,''
  Phys.\ Rev.\ D {\bf 92}, no. 6, 064021 (2015)
  [arXiv:1506.03505 [hep-th]];
  A.~Chaney and A.~Stern,
  ``Fuzzy $CP^2$ spacetimes,''
  Phys.\ Rev.\ D {\bf 95}, no. 4, 046001 (2017)
  [arXiv:1612.01964 [hep-th]].
   
  
\bibitem{Vilenkin:1982de} 
  A.~Vilenkin,
  ``Creation of Universes from Nothing,''
  Phys.\ Lett.\  {\bf 117B}, 25 (1982).
  
\bibitem{Hanada:2014ima}
  M.~Hanada and H.~Shimada,
  ``On the continuity of the commutative limit of the 4d N=4 non-commutative super Yang–Mills theory,''
  Nucl.\ Phys.\ B {\bf 892} (2015) 449
  doi:10.1016/j.nuclphysb.2015.01.016
  [arXiv:1410.4503 [hep-th]].
  
  
    
  \bibitem{Grosse:1996mz} 
  H.~Grosse, C.~Klimcik and P.~Presnajder,
  ``On finite 4-D quantum field theory in noncommutative geometry,''
  Commun.\ Math.\ Phys.\  {\bf 180}, 429 (1996)
  [hep-th/9602115].
  
   \bibitem{Castelino:1997rv} 
  J.~Castelino, S.~Lee and W.~Taylor,
  ``Longitudinal five-branes as four spheres in matrix theory,''
  Nucl.\ Phys.\ B {\bf 526}, 334 (1998)
  [hep-th/9712105].
  
\bibitem{Ramgoolam:2001zx} 
  S.~Ramgoolam,
  ``On spherical harmonics for fuzzy spheres in diverse dimensions,''
  Nucl.\ Phys.\ B {\bf 610}, 461 (2001)
  [hep-th/0105006];
  P.~M.~Ho and S.~Ramgoolam,
  ``Higher dimensional geometries from matrix brane constructions,''
  Nucl.\ Phys.\ B {\bf 627}, 266 (2002)
  [hep-th/0111278].
 
  
\bibitem{Kimura:2002nq} 
  Y.~Kimura,
  ``Noncommutative gauge theory on fuzzy four sphere and matrix model,''
  Nucl.\ Phys.\ B {\bf 637}, 177 (2002)
  [hep-th/0204256].
   
  \bibitem{Medina:2002pc} 
  J.~Medina and D.~O'Connor,
  ``Scalar field theory on fuzzy S**4,''
  JHEP {\bf 0311}, 051 (2003)
  [hep-th/0212170].
  
\bibitem{Sperling:2017dts} 
  M.~Sperling and H.~C.~Steinacker,
  ``Covariant 4-dimensional fuzzy spheres, matrix models and higher spin,''
  J.\ Phys.\ A {\bf 50}, no. 37, 375202 (2017)
  [arXiv:1704.02863 [hep-th]].
  
 \bibitem{Steinacker:2015dra} 
  H.~C.~Steinacker,
  ``One-loop stabilization of the fuzzy four-sphere via softly broken SUSY,''
  JHEP {\bf 1512}, 115 (2015)
  [arXiv:1510.05779 [hep-th]].
  
  \bibitem{Steinacker:2011ix} 
  H.~Steinacker,
  ``Non-commutative geometry and matrix models,''
  PoS QGQGS {\bf 2011}, 004 (2011)
  [arXiv:1109.5521 [hep-th]].
 
  \bibitem{Azuma:2004yg}
  T.~Azuma, S.~Bal, K.~Nagao and J.~Nishimura,
  ``Absence of a fuzzy S**4 phase in the dimensionally reduced 5-D Yang-Mills-Chern-Simons model,''
  JHEP {\bf 0407} (2004) 066
  [hep-th/0405096]. 
  
  
  
  
  \bibitem{Hasebe:2012mz} 
  K.~Hasebe,
  ``Non-Compact Hopf Maps and Fuzzy Ultra-Hyperboloids,''
  Nucl.\ Phys.\ B {\bf 865}, 148 (2012)
  [arXiv:1207.1968 [hep-th]].
  
 \bibitem{Grosse:2010tm} 
  H.~Grosse, P.~Presnajder and Z.~Wang,
  ``Quantum Field Theory on quantized Bergman domain,''
  J.\ Math.\ Phys.\  {\bf 53}, 013508 (2012)
  [arXiv:1005.5723 [math-ph]].
  
  
 \bibitem{Fernando:2009fq}
  S.~Fernando and M.~Gunaydin,
  ``Minimal unitary representation of SU(2,2) and its deformations as massless conformal fields and their supersymmetric extensions,''
  J.\ Math.\ Phys.\  {\bf 51} (2010) 082301
  [arXiv:0908.3624 [hep-th]];
  M.~Gunaydin, D.~Minic and M.~Zagermann,
  ``4D doubleton conformal theories, CPT and IIB string on AdS$_5 \times$ S$^5$,''
  Nucl.\ Phys.\ B {\bf 534}, 96 (1998)
  Erratum: [Nucl.\ Phys.\ B {\bf 538}, 531 (1999)]
  [hep-th/9806042].
  
  
\bibitem{Mack:1975je} 
  G.~Mack,
  ``All Unitary Ray Representations of the Conformal Group SU(2,2) with Positive Energy,''
  Commun.\ Math.\ Phys.\  {\bf 55}, 1 (1977).
  
  
\bibitem{Mack:1969dg} 
  G.~Mack and I.~Todorov,
  ``Irreducibility of the ladder representations of u(2,2) when restricted to the poincare subgroup,''
  J.\ Math.\ Phys.\  {\bf 10}, 2078 (1969).

  
\bibitem{Heidenreich:1980xi} 
  W.~Heidenreich,
  ``Tensor Products of Positive Energy Representations of SO(3,2) and SO(4,2),''
  J.\ Math.\ Phys.\  {\bf 22}, 1566 (1981).

  
 \bibitem{Buric:2017yes} 
  M.~Buric, D.~Latas and L.~Nenadovic,
  ``Fuzzy de Sitter Space,''
  arXiv:1709.05158 [hep-th];
  M.~Buric and J.~Madore,
  ``Noncommutative de Sitter and FRW spaces,''
  Eur.\ Phys.\ J.\ C {\bf 75}, no. 10, 502 (2015)
  [arXiv:1508.06058 [hep-th]].
  
  
\bibitem{Gazeau:2006hj} 
  J.~P.~Gazeau, J.~Mourad and J.~Queva,
  ``Fuzzy de Sitter space-times via coherent states quantization,''
  quant-ph/0610222;
  J.~P.~Gazeau and F.~Toppan,
  ``A Natural fuzzyness of de Sitter space-time,''
  Class.\ Quant.\ Grav.\  {\bf 27}, 025004 (2010)
  [arXiv:0907.0021 [hep-th]].
  
 
\bibitem{Heckman:2014xha} 
  J.~Heckman and H.~Verlinde,
  ``Covariant non-commutative space–time,''
  Nucl.\ Phys.\ B {\bf 894}, 58 (2015)
  [arXiv:1401.1810 [hep-th]].  
  
  
 \bibitem{Klammer:2009ku} 
  D.~Klammer and H.~Steinacker,
  ``Cosmological solutions of emergent noncommutative gravity,''
  Phys.\ Rev.\ Lett.\  {\bf 102}, 221301 (2009)
  [arXiv:0903.0986 [gr-qc]].
  
  
\bibitem{Sperling:2017gmy} 
   M.~Sperling and H.~C.~Steinacker,
  ``Higher spin gauge theory on fuzzy $S^4_N$,''
  arXiv:1707.00885 [hep-th].
  

 \bibitem{Chamseddine:2014nxa} 
  A.~H.~Chamseddine, A.~Connes and V.~Mukhanov,
  ``Quanta of Geometry: Noncommutative Aspects,''
  Phys.\ Rev.\ Lett.\  {\bf 114}, no. 9, 091302 (2015)
  doi:10.1103/PhysRevLett.114.091302
  [arXiv:1409.2471 [hep-th]].   
  
  
  
  
  
  
\end{thebibliography}
\end{document}